\documentclass[physrev,reprint,superscriptaddress,bibnotes]{revtex4-2}  

\usepackage[T1]{fontenc}
\usepackage[utf8]{inputenc} 
\usepackage[australian]{babel}

\usepackage[a4paper,centering,hmargin=1.75cm,vmargin=2cm]{geometry} 
\usepackage{amsmath,amssymb,graphicx,bm,microtype} 
\usepackage[colorlinks,allcolors=blue!50!black]{hyperref} 
\usepackage[all]{hypcap}
\usepackage{cleveref,braket,siunitx}
\usepackage[version=4]{mhchem}
\usepackage{lipsum}

\usepackage{tikz}

\begin{document}

\title{Robust and Deterministic Preparation of Bosonic Logical States in a Trapped Ion}

\author{V.~G.~Matsos}
\affiliation{School of Physics, University of Sydney, NSW 2006, Australia}
\affiliation{ARC Centre of Excellence for Engineered Quantum Systems, University of Sydney, NSW 2006, Australia}

\author{C.~H.~Valahu}
\affiliation{School of Physics, University of Sydney, NSW 2006, Australia}
\affiliation{ARC Centre of Excellence for Engineered Quantum Systems, University of Sydney, NSW 2006, Australia}
\affiliation{Sydney Nano Institute, University of Sydney, NSW 2006, Australia}

\author{T.~Navickas}
\affiliation{School of Physics, University of Sydney, NSW 2006, Australia}
\affiliation{ARC Centre of Excellence for Engineered Quantum Systems, University of Sydney, NSW 2006, Australia}

\author{A.~D.~Rao}
\affiliation{School of Physics, University of Sydney, NSW 2006, Australia}
\affiliation{ARC Centre of Excellence for Engineered Quantum Systems, University of Sydney, NSW 2006, Australia}

\author{M.~J.~Millican}
\affiliation{School of Physics, University of Sydney, NSW 2006, Australia}
\affiliation{ARC Centre of Excellence for Engineered Quantum Systems, University of Sydney, NSW 2006, Australia}

\author{X.~C.~Kolesnikow}
\affiliation{School of Physics, University of Sydney, NSW 2006, Australia}
\affiliation{ARC Centre of Excellence for Engineered Quantum Systems, University of Sydney, NSW 2006, Australia}

\author{M.~J.~Biercuk}
\affiliation{School of Physics, University of Sydney, NSW 2006, Australia}
\affiliation{ARC Centre of Excellence for Engineered Quantum Systems, University of Sydney, NSW 2006, Australia}

\author{T.~R.~Tan}
\email{tingrei.tan@sydney.edu.au}
\affiliation{School of Physics, University of Sydney, NSW 2006, Australia}
\affiliation{ARC Centre of Excellence for Engineered Quantum Systems, University of Sydney, NSW 2006, Australia}
\affiliation{Sydney Nano Institute, University of Sydney, NSW 2006, Australia}

\begin{abstract}
Encoding logical qubits in bosonic modes provides a potentially hardware-efficient implementation of fault-tolerant quantum information processing. Here, we demonstrate high-fidelity and deterministic preparation of highly non-classical bosonic states in the mechanical motion of a trapped ion. Our approach implements error-suppressing pulses through optimized dynamical modulation of laser-driven spin-motion interactions to generate the target state in a single step. We demonstrate logical fidelities for the Gottesman-Kitaev-Preskill (GKP) state as high as $\bar{\mathcal{F}}=0.940(8)$, a distance-3 binomial state with an average fidelity of $\mathcal{F}=0.807(7)$, and a 12.91(5) dB squeezed vacuum state.
\end{abstract}

\maketitle

Fault-tolerant quantum error correction (QEC) for quantum information processing (QIP) necessitates the implementation of a redundant encoding within a sufficiently large Hilbert space to yield protection against local hardware errors~\cite{nielsen2000}. At present, the predominant strategy focuses on encoding a logical qubit with multiple discrete-variable physical qubits manipulated with ultra-low operational errors~\cite{Raussendorf2007, Fowler2012}. 
This approach is resource-intensive, and, despite many impressive demonstrations~\cite{RyanAnderson2021,Egan2021,Zhao2022,Krinner2022,Chen2022,Google2023}, the viability of using QEC in this way to deliver net improvements in hardware error rates remains challenging. 
Many analyses indicate that a large ratio of physical-to-logical qubits is necessary for fault-tolerant operation in target algorithms, posing a substantial resource penalty and far outstripping device sizes available in the near future~\cite{Watson2014}.
An alternative approach involves encoding logical qubits within continuous-variable systems~\cite{Lloyd1999,Braunstein2005}. In particular, the infinite-dimensional Hilbert space spanned by the bosonic mode of a harmonic oscillator offers a highly symmetrical physical system that lends itself to logical encodings including Gottesman-Kitaev-Preskill (GKP)~\cite{GKP2001}, binomial~\cite{Michael2016}, and cat~\cite{Mirrahimi2014} codes. This approach demands fewer individual physical devices at the cost of increased complexity in preparing and controlling the logical code words. 

\begin{figure}[t!]
    \centering
    \includegraphics[]{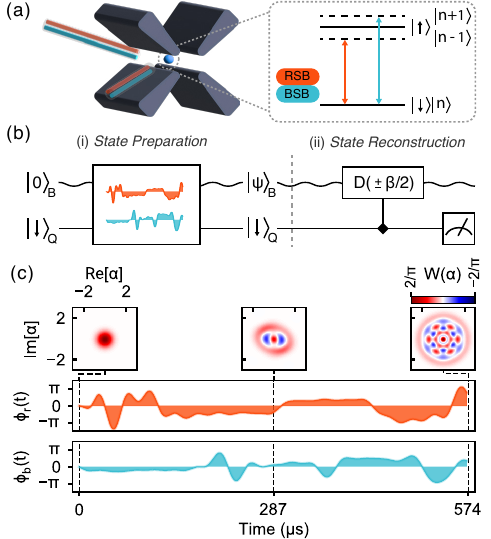}
    \caption{\textbf{State-preparation of non-classical bosonic states in an ion trap.} 
    \textbf{a)} Two orthogonal Raman beams couple the spin and motion of a trapped ion, inducing spin-dependent forces. \textbf{b)} Experimental pulse sequence: (i) the bosonic mode and qubit are initialized to their ground state, the control pulse evolves the system to the target bosonic state under $\hat{H}(t)$ (Eq.~\ref{rsb+bsb}). (ii) Reconstruction of characteristic function, $\chi(\beta)$, by applying a displacement, $\hat{\mathcal{D}}(\pm \beta/2)$, conditioned on the qubit state in the $\hat{\sigma}_x$ basis. Readout of the ancilla qubit in the $\hat{\sigma}_z$ basis measures $\mathrm{Re}[\chi(\beta)]$. \textbf{c)} Target bosonic states are prepared by modulating $\phi_{r, b}(t)$ of the bichromatic fields. Insets show the time-evolution of the Wigner function for the $(|0\rangle+\sqrt{3}|6\rangle) / 2$ binomial state.} 
    \label{fig:GPExp}
\end{figure}

Several experimental works have successfully created different classes of bosonic logical states~\cite{Flhmann2019,Wang2019,McCormick2019,Eickbusch2022,Kudra2022, konno2024}, implemented logical gate sets \cite{Heeres2016}, and demonstrated QEC protocols~\cite{Ofek2016,Hu2019,Campagne-Ibarcq2020,Neeve2022,Sivak2023, Ni2023,Lachance2023}. However, preparing high-quality bosonic codes for use in QEC remains a significant challenge. For instance, attaining fault tolerance by concatenating the GKP code with discrete-variable error-correcting codes requires a squeezing parameter currently estimated at approximately 10 dB~\cite{Menicucci2014,Fukui2018}, a threshold yet to be experimentally reached. Additionally, only the lowest-order binomial code words (distance-2) have been experimentally realized so far, while a minimum distance of 3 is necessary to correct all types of bosonic errors~\cite{Michael2016}.

\begin{figure*}[t]
    \centering
   \includegraphics[]{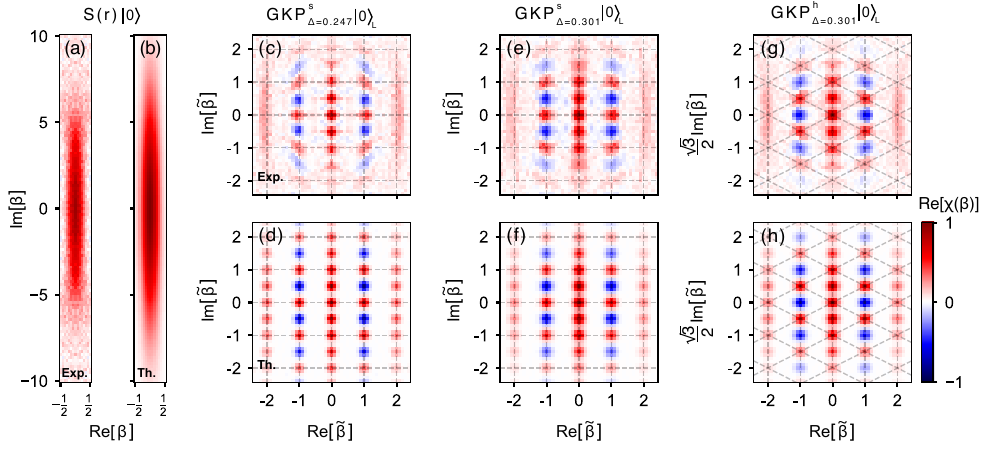}
    \caption{\textbf{Experimentally reconstructed characteristic functions of squeezed and GKP states.} \textbf{(a-b)}~Squeezed state with a target squeezing parameter $r=1.55$ created with only coherent first-order sideband interactions. 
    \textbf{(c-f)}~Approximate square GKP states with target envelope parameters of $\Delta=0.247$ and $\Delta=0.301$, respectively. 
    \textbf{(g-h)}~Approximate hexagonal GKP state with target $\Delta=0.301$. 
    The theoretical characteristic functions of each GKP state are plotted below each sub-figure. The phase space coordinates are normalized with $\tilde{\beta} =\beta / \sqrt{2 \pi}$ for the square and $\tilde{\beta} = \beta / \sqrt{\sqrt{3} \pi}$ for the hexagonal GKP plots. 
    The achieved fidelities, squeezing levels, and pulse durations are summarized in Table \ref{fidelity_table}. 
    The theoretical decoherence-free fidelities are $\mathcal{F}_\mathrm{th.} \geq 0.99$ for the squeezed state and $\mathcal{F}_\mathrm{th.} \geq 0.975$ for the GKP states. 
    We attribute the circular smearing in (c, e) to motional mode dephasing. 
    }
    \label{fig:experimentdata}
\end{figure*}

In this work, we experimentally demonstrate high-fidelity deterministic preparation of a variety of bosonic states by integrating concepts of error suppression via robust control into the protocol for QEC encoding. Code word generation is achieved in a single step using optimized time-domain modulation of the control fields used to manipulate the qubit and motional modes of a trapped ion. The control pulses, obtained from a numerical optimizer, are designed to exhibit robustness against motional dephasing~\cite{boulder_opal1}. We demonstrate the versatility of our approach by creating several classes of bosonic states: a highly squeezed \SI{12.91(5)}{dB} state; square and hexagonal GKP states with a logical fidelity and squeezing as high as $0.940(8)$ and $\SI{7.5(2)}{dB}$, respectively; and the first realization of a distance-3 binomial logical state that can simultaneously protect against universal (correctable) bosonic errors. Finally, we demonstrate the robustness of our protocol to systematic imperfections by demonstrating up to $4.8\times$ lower error than non-robust solutions in the presence of quasistatic motional-frequency offsets.

State-preparation is implemented by laser-induced spin-boson interactions for a single trapped ion described by the effective Hamiltonian
\begin{align}
\label{rsb+bsb}
\hat{H}(t) = \frac{\Omega_r (t)}{2} \hat{\sigma}^{+} \hat{a} e^{i \phi_r(t)} + \frac{\Omega_b (t)}{2} \hat{\sigma}^{+} \hat{a}^{\dagger} e^{i \phi_b(t)} +\text{h.c.},
\end{align}
where $\hat{\sigma}^+ = \ket{\uparrow}\bra{\downarrow}$ is the raising operator of the qubit and $\hat{a}$ ($\hat{a}^\dagger$) is the annihilation (creation) operator of the oscillator. The two terms in the Hamiltonian are referred to as the red- (RSB) and blue-sideband (BSB) interactions~\cite{Wineland1998} with their respective time-dependent Rabi rates, $\Omega_{r,b}(t)$, and phases, $\phi_{r,b}(t)$. This Hamiltonian provides sufficient control to produce the bosonic states considered in this work, where the required nonlinearities for these states are generated from non-commuting terms in the Hamiltonian at different times (see SM). Furthermore, universal control of the qubit-oscillator system may be obtained through the addition of a carrier interaction $\hat{H}_c(t) = \Omega_c (t) \hat{\sigma}^+ e^{i\phi_c(t)}/2 + \textrm{h.c.}$~\cite{Law1996,Eickbusch2022}, but is not necessary to prepare the states considered in this Letter.

We numerically design the time-dependent controls of $\hat{H}(t)$ such that a target bosonic state is created via the time evolution operator $e^{-i \int \hat{H}(t)dt}$ with $\{\Omega_{r,b}(t),\phi_{r,b}(t)\}$ as optimizable control parameters. Optimized control pulses are obtained through a gradient-based constrained optimization using Q-CTRL's Boulder Opal package \cite{boulder_opal1,boulder_opal2}. 
Constraints are included on candidate control pulses to improve experimental implementation: first, the Rabi rates of the RSB and BSB interactions are kept constant throughout the evolution ($\Omega_r(t) = \Omega_b(t) = 2\pi\times\SI{2}{kHz}$) to avoid unwanted time-varying Stark shifts; second, slew-rate and filtering constraints are added to the optimized waveform phases to comply with hardware limitations. We define the cost function to maximize pulse fidelity for a target state-preparation task in the presence of motional dephasing (the dominant source of noise in our system) while also minimizing total pulse durations (see SM). 

Our experiment is performed with a single \ce{^{171}Yb^+} ion confined in a macroscopic Paul trap, with radial frequencies $\{\omega_x, \omega_y\} = 2\pi \times \{1.33, 1.51\}$~MHz. The bosonic states are encoded in the radial-$x$ mode, while the ancillary qubit is made up of the two magnetically insensitive hyperfine levels of the \ce{^2S_{1/2}} ground state with $\ket{\downarrow} = \ket{F=0, m_F=0}$ and $\ket{\uparrow} = \ket{F=1, m_F = 0}$. Coherent spin-motion interactions are enacted by a $\SI{355}{nm}$ pulsed laser through stimulated Raman transitions. Phase modulation of the RSB and BSB interactions is implemented by adjusting a radio-frequency (RF) signal that drives an acousto-optic-modulator (AOM) in the path of the Raman beam (see Refs.~\cite{MacDonell2023, Valahu2023} for more details on the experimental system).
\begin{figure}[t]
    \centering
   \includegraphics[]{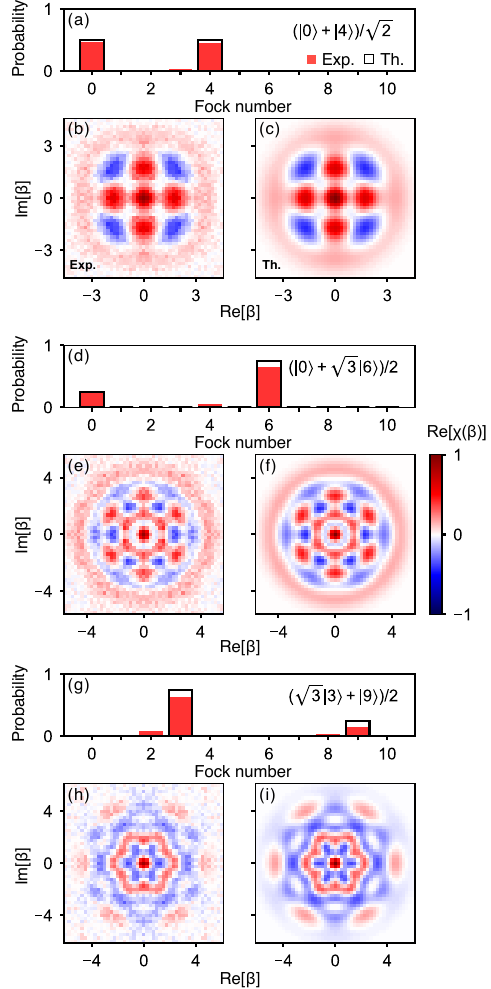}
    \caption{\textbf{Reconstructed characteristic function of binomial code words.} 
    Experimental and theoretical characteristic functions of 
    \textbf{(b, c)} $\ket{{+Z}_{S=1}}_\mathrm{bin} = \frac{1}{\sqrt{2}}(\ket{0}+\ket{4})$, 
    \textbf{(e, f)} $\ket{{+Z}_{S=2}}_\mathrm{bin} = \frac{1}{2}(\ket{0}+\sqrt{3}\ket{6})$ and 
    \textbf{(h, i)} $\ket{{-Z}_{S=2}}_\mathrm{bin} = \frac{1}{2}(\sqrt{3}\ket{3}+\ket{9})$. 
    \textbf{(a, d, g)} Theoretical (unshaded, black) and experimental (shaded, red) Fock number occupation probabilities, calculated from $\hat{\rho}_\mathrm{exp}$. Preparation of $\ket{{-Z}_{S=2}}_\mathrm{bin}$ begins with the ancilla qubit in the $\ket{\uparrow}$ state. The theoretical decoherence-free fidelities for each state are $\mathcal{F}_\mathrm{th.} \geq 0.9999$.
    }
    \label{fig:binomialdata}
\end{figure}

The experimental pulse sequence used in \textit{state-preparation} is shown in Fig.~\ref{fig:GPExp}; after initializing the qubit and bosonic mode to their ground state, $\ket{\downarrow}\otimes\ket{0}$, we apply the Hamiltonian of Eq.~\ref{rsb+bsb} with the numerically optimized pulses to prepare the target state $\ket{\downarrow}\otimes\ket{\psi}$. This is followed by a \textit{state reconstruction} protocol that aims to retrieve the density matrix of the experimentally generated state. To this end, we measure the characteristic function 
\begin{equation}
\chi(\beta)=\langle\hat{\mathcal{D}}(\beta)\rangle,
\end{equation}
where $\hat{\mathcal{D}}(\beta)=e^{\beta \hat{a}^{\dagger}-\beta^* \hat{a}}$ is the displacement operator and $\beta \in \mathbb{C}$ is a point in phase space. 

The characteristic function is sampled using the protocol outlined in~\cite{Flhmann2020}. We apply a state-dependent force (SDF) $\hat{\mathcal{D}}(\beta(t) \hat{\sigma}_x / 2)$, where $\beta(t) / 2=\Omega t e^{-i \left(\phi_r-\phi_b\right) / 2}/2$, which maps information from the bosonic mode onto the qubit for readout. The SDF is implemented by simultaneously applying the RSB and BSB interactions with $\Omega_r = \Omega_b = \Omega$. The real part of the characteristic function, $\mathrm{Re}[\chi(\beta)]$, is determined through projective readout of the internal qubit state of the ion (the imaginary part of the characteristic function can be obtained by applying a qubit rotation prior to the SDF). We only perform measurements in the positive quadrant of phase space. For the states considered in this work, the remaining quadrants are obtained in post-processing by the symmetry of $\chi(\beta)$ with respect to both phase space axes. Furthermore, we only reconstruct $\mathrm{Re}[\chi(\beta)]$ as the imaginary part vanishes under this same symmetry.

\renewcommand{\arraystretch}{1.3}
\begin{table}[t]
\small
\addtolength{\tabcolsep}{-2pt}
\begin{tabular}{cccc}
\hline \hline \text {State}  & \text {Fidelity}& \text { Squeezing (dB) } & \text{Time (}\text{\SI{}{\micro\second})}\\
\hline $\hat{S}(r=1.55)\ket{0}$  & 0.753(4) & 12.91(5) & 1057\\
$\mathrm{GKP}^\mathrm{s}_{\Delta = 0.247}$ & $0.90(1)_\mathrm{L}, 0.60(1)$ &  5.5(2), 6.3(2) & 1301\\
$\mathrm{GKP}^\mathrm{s}_{\Delta = 0.301}$ & $0.940(8)_\mathrm{L}, 0.83(1)$ & 7.5(2), 7.5(2)  & 933\\
$\mathrm{GKP}^\mathrm{h}_{\Delta = 0.301}$ & $0.91(1)_\mathrm{L}, 0.77(3)$ & 6.5(3), 6.3(4) & 978\\
$(\ket{0}+\ket{4})/\sqrt{2}$  & 0.889(9) & - & 514\\
$(\ket{0}+\sqrt{3}\ket{6})/2 $  & 0.843(9) & - &688\\
$(\sqrt{3}\ket{3}+\ket{9})/2 $  & 0.77(1) & - &780\\
\hline \hline
\end{tabular}
\label{fidelity_table}
\caption{\textbf{Fidelities, squeezings and durations of experimentally prepared bosonic states.} The state-preparation fidelities, defined as $\mathcal{F}= \bra{\psi_t}\hat{\rho}_\mathrm{exp}\ket{\psi_t} $, are calculated from the reconstructed density matrices $\hat{\rho}_\mathrm{exp}$ of the data in Fig.~\ref{fig:experimentdata} and Fig.~\ref{fig:binomialdata}. We also report the logical fidelities of the GKP states, denoted by the subscript L. 
Uncertainties correspond to the 1-sigma deviation obtained from bootstrapping. 
}
\end{table}

We experimentally generate several bosonic states to demonstrate the versatility of our protocol. Each state is prepared by implementing a targeted control solution obtained from the numerical optimizer with results summarized in Table~\ref{fidelity_table}. The density matrix, $\hat{\rho}_\mathrm{exp}$, is retrieved from the experimentally measured characteristic function using a convex optimization procedure~\cite{Ahmed2021}~(see SM). State-preparation fidelity is then computed as $\mathcal{F}=\left\langle\psi_t\left|\hat{\rho}_\mathrm{exp}\right| \psi_t\right\rangle$, where $\ket{\psi_t}$ is the target state. For GKP states, we also report the logical fidelity that quantifies the amount of logical information contained in the state~\cite{Kolesnikow2024}. We bootstrap these measurements to determine the associated uncertainties~\cite{Efron1994}.

We first prepare a squeezed state $\hat{S}(r) \ket{0}$, where $\hat{S}(r) = \exp{\big[ \frac{1}{2} \left(r^*\hat{a}^2 -r \hat{a}^{\dagger2}\right)  } \big]$, with a target squeezing parameter of $r = 1.55$. The reconstructed characteristic function is shown in Fig.~\ref{fig:experimentdata}(a), which we fit to theory \cite{Flhmann2020} and find a squeezing parameter $r=1.487(5)$ ($\SI{12.91(5)}{dB}$). We detail squeezing estimations using multiple analysis methods in the SM. We determine a fidelity of $0.753(4)$, where the accuracy is limited by motional dephasing during reconstruction. 

We next prepare approximate GKP code words (see SM for a detailed description). The exact GKP codespace is defined as the mutual +1 eigenspace of the stabilizer displacement operators $\hat{{S}}_X = \hat{\mathcal{D}}(2\alpha)$ and $\hat{{S}}_Z = \hat{\mathcal{D}} (2\beta)$, for $\beta \alpha^*-\beta^* \alpha=  i \pi$ and $\alpha,\beta \in \mathbb{C}$. This definition admits unphysical GKP code words whose Wigner functions are two-dimensional grids of delta functions with infinite extent. GKP states can be made physical by applying a Gaussian envelope characterized by the parameter $\Delta \in [0,1]$, where the exact code words are recovered in the limit $\Delta \rightarrow 0$~\cite{GKP2001}.  

We prepare the $\ket{0}_\mathrm{L}$ logical states of a square and a hexagonal GKP code with target squeezings of $\SI{10.43}{dB}$ ($\Delta = 0.301$), and an additional square GKP code with $\SI{12.15}{dB}$ ($\Delta = 0.247$) (see Fig.~\ref{fig:experimentdata}(c)-(h)). The preparation of $\ket{1}_\mathrm{L}$ states is not shown experimentally; however, our pulse optimization scheme produces similar results in fidelity and duration. The state-preparation fidelity, $\mathcal{F}$, and \textit{logical} fidelity, $\bar{\mathcal{F}}$ (see SM), are reported in Table~\ref{fidelity_table}. 

The achieved squeezings of the GKP states are calculated from $\hat{\rho}_\mathrm{exp}$ with $\langle\hat{S}_X\rangle$ and $\langle\hat{S}_Z\rangle$, as defined in Ref.~\cite{Duivenvoorden2017} (see Table~\ref{fidelity_table}). Alternatively, using the analysis method of Ref.~\cite{Eickbusch2022} results in squeezings of \{GKP$^{s}_{\Delta=0.247}$, GKP$^{s}_{\Delta=0.301}$, GKP$^{h}_{\Delta=0.301}$\} = \{10.4(1), 9.47(8), 9.1(1)\} dB, which are considerably higher than the values calculated above. The discrepancy in the results obtained by the two methods motivates a more comprehensive characterization method for GKP states (a discussion on these methods is included in the SM). 

We also prepare binomial code words, which encode logical qubits in finite superpositions of Fock states~\cite{Michael2016} (see SM for a detailed description). Binomial codes can exactly correct for errors that are polynomial in creation and annihilation operators \cite{Marios2016}. We first create the state $\ket{{+Z}_{S=1}}_\mathrm{bin} = \frac{1}{\sqrt{2}}(\ket{0}+\ket{4})$ [see Fig.~\ref{fig:binomialdata}(a-c)], which can protect against errors linear in $\hat{a}$ or $\hat{a}^\dagger$. 
We then create distance-3 binomial states, $\ket{{+Z}_{S=2}}_\mathrm{bin} = \frac{1}{2}(\ket{0}+\sqrt{3}\ket{6})$ and $\ket{{-Z}_{S=2}}_\mathrm{bin} = \frac{1}{2}(\sqrt{3}\ket{3}+\ket{9})$ [see Fig~\ref{fig:binomialdata}(d-i)], which protect against errors quadratic in $\hat{a}$ or $\hat{a}^\dagger$~\cite{Michael2016}. The achieved fidelities are reported in Table~\ref{fidelity_table}. 
To the best of our knowledge, this work appears to be the first experimental preparation of distance-3 binomial code words.

\begin{figure}[t]
    \centering
   \includegraphics[]{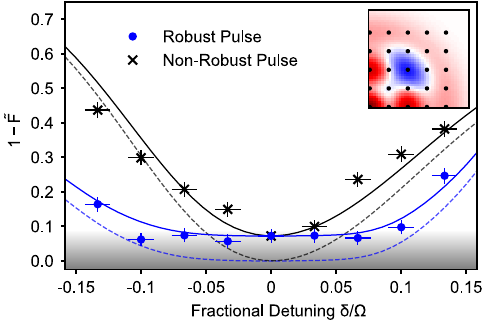}
    \caption{\textbf{Robustness against errors in motional frequency in preparing a binomial state.}   Experimental offsets are engineered in software by adding symmetric frequency shifts to the red- and blue-sideband fields.
    Pseudo-fidelity measurements indicate that the \textit{robust} pulse (blue, $T=\SI{1073}{\mu s}$) is less sensitive to noise than the \textit{non-robust} pulse (black, $T=\SI{759}{\mu s}$). 
    The inset shows the top right quadrant of the $(\ket{0} + \ket{4})/\sqrt{2}$ binomial state's characteristic function, and black markers indicate the $5\times 5$ points that were sampled to measure $\tilde{\mathcal{F}}$ (Eq.~\ref{eq:pseudo_fidelity}).
    The dashed theory line is obtained from numerical simulations.
    The solid theory line includes a scaling correction to account for infidelities during preparation and reconstruction to indicate the qualitative agreements between SPAM-adjusted theoretical results and experiment. The gray shaded region depicts the SPAM error floor, which we define by the fidelity of the robust pulse at $\delta = 0$. The $y$ error bars indicate the uncertainty from quantum projection noise (500 repetitions). The $x$ error bars are the uncertainty of the applied frequency offset due to drifts of the motional mode.
    }
    \label{fig:pseudo_infidelity}
\end{figure}

Finally, we examine the impact of incorporating robustness against motional dephasing by comparing  \textit{robust} and \textit{non-robust} optimized state-preparation protocols. Here, we remove duration constraints in the optimization for both pulses, and 
the \textit{non-robust} pulse is optimized without $\hat{a}^\dagger \hat{a}$ Hamiltonian terms (see SM). Experimentally, we compare both protocols for binomial state $\ket{{+Z}_{S=1}}_\mathrm{bin}$ preparation (see Fig.~\ref{fig:pseudo_infidelity}) and measure a pseudo-fidelity 
\begin{equation}
    \label{eq:pseudo_fidelity}
    \tilde{\mathcal{F}} = \frac{1}{\mathcal{N}}\sum_{i} \chi_\textrm{exp}(\beta_i) \chi_\textrm{th}(\beta_i),
\end{equation}
which computes the overlap of experimental and theoretical characteristic functions \cite{Rundle2021}. Here, $\mathcal{N} = 1/\sum_i \chi_\mathrm{th}(\beta_i)^2$ is a normalization factor. This strategy provides a qualitative comparison with fewer measurements. We quantify the robustness in the presence of applied quasistatic $\delta \hat{a}^\dagger \hat{a}$ offsets, which can be associated with miscalibrations of the motional frequency. Figure~\ref{fig:pseudo_infidelity} illustrates that the fidelity of the state generated using a \textit{robust} pulse remains stable despite errors up to $\delta/\Omega = \pm 0.1$, while errors accumulate rapidly for the \textit{non-robust} pulse as $|\delta|$ increases from zero. Furthermore, the \textit{robust} pulse achieves up to 4.8$\times$ lower error for $\delta/\Omega = - 0.1$. Measurements for both cases align well with theoretical predictions when incorporating SPAM (state-preparation and measurement) errors.

The main error mechanisms affecting achieved fidelities and squeezing parameters are thermal noise and motional dephasing (see SM). Thermal noise results from imperfect cooling, with typically achieved average phonon occupancies of $\bar{n} = 0.05-0.1$. Motional dephasing noise arises from noise in the ion trap's RF resonator and its electronic circuit. Numerical simulations suggest that dephasing also leads to significant infidelities during the reconstruction protocol, potentially dominating the results reported in Table~\ref{fidelity_table}. Approximately 3-7\% of the measured characteristic function error also results from residual spin-motion entanglement, causing $\mathrm{Re}[\chi(0)] \neq 1$ after the pulse. This can be mitigated by a mid-circuit measurement before the reconstruction pulse; we have opted against this due to substantial motional dephasing-induced errors during measurement.

In summary, we have demonstrated a deterministic protocol for generating a wide range of high-fidelity bosonic states relevant to QEC using numerically optimized and error-robust pulses. Our scheme only requires phase modulation of first-order red- and blue-sideband transitions, compatible with conventional experimental techniques. This approach surpasses gate-based methods \cite{Hastrup2021, Neeve2022} as pulse-level optimization enables higher theoretical fidelities for a given duration (see SM).

The protocol's versatility, robustness to dephasing, and the quality of the states demonstrated in this work suggest that this method is promising for bosonic QEC. Our achieved squeezed state only uses first-order sideband interactions and offers an alternative to previous trapped-ion demonstrations~\cite{Meekhof1996, Kienzler2015, Burd2019}. Furthermore, the distance-3 binomial code words created by our protocol demonstrate the experimental feasibility of implementing logical codes that can simultaneously protect against boson loss, gain, and dephasing events \cite{Michael2016}. Our GKP states show high logical fidelities and squeezings and may also be useful for quantum sensing of displacements \cite{Duivenvoorden2017}, opening new opportunities in precision measurement. 

\begin{acknowledgments}
We thank Thomas Smith for helpful discussions surrounding GKP and binomial codes. We were supported by the U.S. Office of Naval Research Global (N62909-20-1-2047), the U.S. Army Research Office Laboratory for Physical Sciences (W911NF-21-1-0003), the U.S. Air Force Office of Scientific Research (FA2386-23-1-4062), the U.S. Intelligence Advanced Research Projects Activity (W911NF-16-1-0070), Lockheed Martin, the Sydney Quantum Academy (ADR, MJM, VGM, and TRT), the University of Sydney Postgraduate Award scholarship (VGM), the Australian Research Council, and H.\ and A.\ Harley.
\end{acknowledgments}

\section*{Data availability}

The experimental data are available in an online repository at https://doi.org/10.5281/zenodo.12741298.


%

\renewcommand{\thefigure}{S\arabic{figure}}
\setcounter{figure}{0}

\renewcommand{\thetable}{S\Roman{table}}
\setcounter{table}{0}

\renewcommand{\theequation}{S\arabic{equation}}
\setcounter{equation}{0}

\title{Supplemental Material: Robust and Deterministic Preparation of Bosonic Logical States in a Trapped Ion}

\maketitle

\subsection{Bosonic Control}

\begin{figure*}[t]
    \centering
    \includegraphics[]{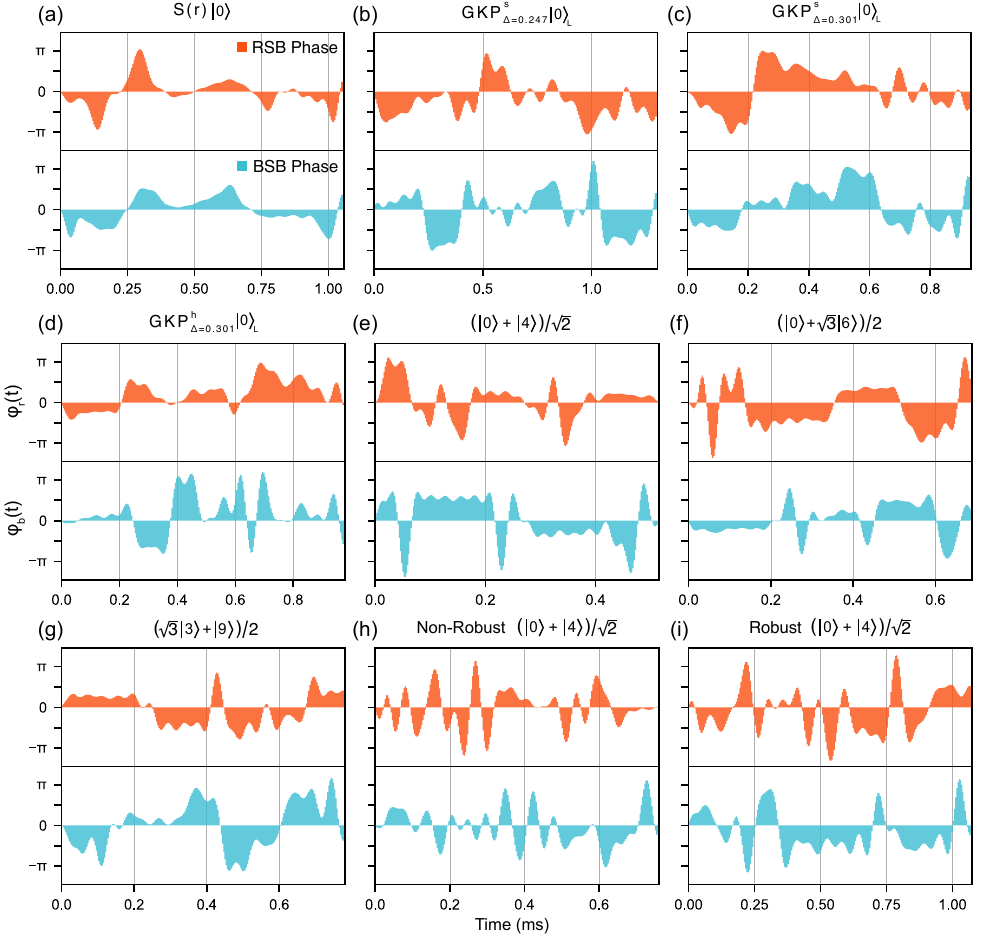}
    \caption{Numerically optimized red-sideband phases $\phi_{r}(t)$ (red) and blue-sideband phases $\phi_b(t)$ (blue) for the state preparation pulses featured in this Letter. Using the Hamiltonian from Eq. 1 and $\Omega_{r,b} = 2\pi \times \SI{2}{kHz}$, each waveform is used to produce the target bosonic states denoted by: \textbf{(a)} Squeezed state $S(r=1.55) \ket{0}$,
    \textbf{(b)} Square GKP state, $\Delta = 0.247$,
    \textbf{(c)} Square GKP state, $\Delta = 0.301$,
    \textbf{(d)} Hexagonal GKP state, $\Delta = 0.301$,
    \textbf{(e)} $\left(\ket{0}+\ket{4}\right)/\sqrt{2}$ Binomial state,
    \textbf{(f)} $\left(\ket{0}+\sqrt{3}\ket{6}\right)/2$ Binomial state,
    \textbf{(g)} $\left(\sqrt{3}\ket{3}+\ket{9}\right)/2$ Binomial state,
    \textbf{(h)} Non-Robust $\left(\ket{0}+\ket{4}\right)/\sqrt{2}$ Binomial state,
    \textbf{(i)} Robust $\left(\ket{0}+\ket{4}\right)/\sqrt{2}$ Binomial state.}
    \label{fig:pulse_plots}
\end{figure*}

Here, we show that the Hamiltonian of Eq.~1 provides sufficient control to generate the bosonic states featured in this Letter.

All bosonic states produced in this work are eigenstates of the parity operator $e^{\mathrm{i} \pi \hat{a}^\dagger \hat{a}}$, with all states having even parity except for the $\ket{{-Z}_{S=2}}_\mathrm{bin}$ binomial state which has odd parity. 
In other words, they are made up of a superposition of only even (odd) Fock states. 
For a bosonic mode initialized to a vacuum state, these states can be generated from Hamiltonians that are even (odd) polynomials in the quadratures $\hat{x} = (\hat{a} + \hat{a}^\dagger)/\sqrt{2}$ and $\hat{p} = -i(\hat{a} - \hat{a}^\dagger)/\sqrt{2}$. 

We now consider a qubit coupled to a bosonic mode. 
Given the initial state $\ket{\downarrow}\otimes\ket{0}$, a state with bosonic even-parity $
\ket{\downarrow}\otimes\ket{\psi_{\mathrm{even}}}$ may be prepared from a unitary transformation generated by Hamiltonians of the form $ \hat{\sigma}_z^k\hat{x}^i\hat{p}^j  + \text{h.c.}$, where $i+j = n$, $n$ is an even integer and $k\in \{0,1\}$. On the other hand, given the initial state $\ket{\uparrow} \otimes \ket{0}$, a state with bosonic odd-parity $\ket{\downarrow}\otimes\ket{\psi_\mathrm{odd}}$ may be prepared by generators of the form $\hat{\sigma}_{x/y}\hat{x}^i\hat{p}^j + \text{h.c.}$, where $i+j = n+1$. 

Here, $\hat{\sigma}_{x/y/z}$ are the Pauli operators for the qubit.

To show that these spin-boson unitary transformations may be generated from our control Hamiltonian, we closely follow the derivations outlined in Ref.~\cite{Eickbusch2022}. For Rabi rates $\Omega_r = \Omega_b = \Omega$, the control Hamiltonian may be expressed in terms of Pauli and quadrature operators as
\begin{equation}
\hat{H} = \dfrac{\Omega}{\sqrt{2}} \bigl(\cos{(\phi_s)}\hat{\sigma}_x+\sin{(\phi_s)}\hat{\sigma}_y\bigr)\bigl(\cos{(\phi_m)}\hat{x}-\sin{(\phi_m)}\hat{p}\bigr),
\end{equation}
where we use the parameterization of phases $\phi_s = (\phi_r+\phi_b)/2$, $\phi_m = (\phi_r -\phi_b)/2$ which relate to modulated laser phases $\phi_r,\phi_b$. 

Hence, by choosing $\phi_s,\phi_m \in \{0, \pi/2\}$, we obtain a set of generating Hamiltonians, $S = \{\hat{\sigma}_x\hat{x}, \hat{\sigma}_y\hat{x}, \hat{\sigma}_x\hat{p}, \hat{\sigma}_y\hat{p}\}$. 
This set can be expanded by making use of the following two identities \cite{Braunstein2005}
\begin{subequations}
\begin{align}
&\mathrm{e}^{-\mathrm{i} \hat{A} \delta t} \mathrm{e}^{-\mathrm{i} \hat{B} \delta t} \mathrm{e}^{\mathrm{i} \hat{A} \delta t} \mathrm{e}^{\mathrm{i} \hat{B} \delta t}=\mathrm{e}^{-[\hat{A}, \hat{B}] \delta t^2}+O\left(\delta t^3\right), \label{eq:Trotter1}\\
&\mathrm{e}^{\mathrm{i} \hat{A} \delta t / 2} \mathrm{e}^{\mathrm{i} \hat{B} \delta t / 2} \mathrm{e}^{\mathrm{i} \hat{B} \delta t / 2} \mathrm{e}^{\mathrm{i} \hat{A} \delta t/2}=\mathrm{e}^{\mathrm{i}(\hat{A}+\hat{B}) \delta t}+O\left(\delta t^3\right),\label{eq:Trotter2}
\end{align}
\end{subequations}
where $\hat{A}$ and $\hat{B}$ are any two operators and $\delta t$ is discretized time. Using $\hat{A},\hat{B}\in S$,
we may expand $S$ to include Hamiltonian terms $\{\hat{\sigma}_z\hat{x}^i \hat{p}^j \}$, where $i + j =2$, by using \cref{eq:Trotter1} together with the commutation relations $[\hat{\sigma}_x \hat{x} , \hat{\sigma}_y\hat{x}]= 2\mathrm{i}\hat{\sigma}_z \hat{x}^2 $, $[\hat{\sigma}_x\hat{p}, \hat{\sigma}_y\hat{p}]= 2\mathrm{i}\hat{\sigma}_z \hat{p}^2 $ and $[\hat{\sigma}_x\hat{x},\hat{\sigma}_y\hat{p}] =  \mathrm{i}(\hat{\sigma}_z\hat{x}\hat{p} + \text{h.c.})$. The set can then be further expanded by iterating these commutators over $S$ to obtain the terms $\hat{\sigma}_z^k\hat{x}^i \hat{p}^j$, where $i+j = n$ and $k\in \{0,1\}$. Furthermore, we can obtain the terms $\hat{\sigma}_x \hat{x}^i \hat{p}^j$ or $ \hat{\sigma}_y\hat{x}^i \hat{p}^j$ where $i+j =n+ 1$ by making use of the commutation relations of these even-powered terms with the Hamiltonians in the original generating set. 

Therefore, together with \cref{eq:Trotter2}, we can produce arbitrary even (odd) polynomials in the bosonic quadratures that preserve (invert) the initial state of the qubit in the $\hat{\sigma}_z$ basis. Thus, the phase-modulated control Hamiltonian is capable of generating the bosonic states in this Letter.

\begin{figure*}[t]
    \centering
    \includegraphics[]{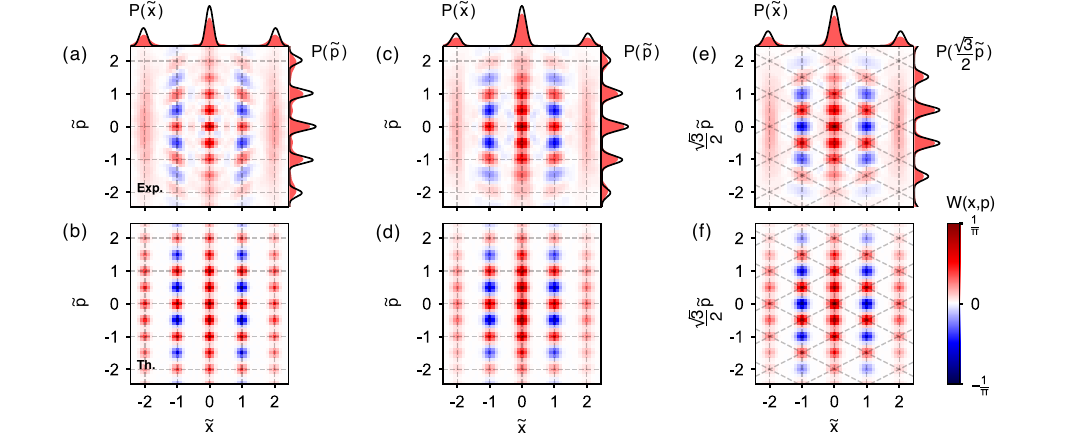}
    \caption{\textbf{Experimentally reconstructed Wigner functions of GKP states.}
    \textbf{(a, b)} Approximate GKP states with targeted envelope parameters $\Delta = 0.247$ and \textbf{(c, d)} $\Delta = 0.301$. 
    \textbf{(e, f)} Approximate hexagonal GKP state with target $\Delta = 0.301$.
    The experimental Wigner functions \textbf{(a, c, e)} are compared to theory \textbf{(b, d, f)}, and are calculated from the density matrices that are reconstructed from the characteristic functions of Fig.~2.
    The phase space coordinates are normalized with $\tilde{x} = x/\sqrt{\pi}$ and $\tilde{p} = p/\sqrt{\pi}$ for the square GKP states, and $\tilde{x} =x\sqrt{2/(\sqrt{3}\pi)} $ and $\tilde{p} = p\sqrt{2/(\sqrt{3}\pi)} $ for the hexagonal GKP state.
    Marginals plot the experimental (red, shaded) and theoretical (black, unshaded) probability distributions along $\tilde{x}$ and $\tilde{p}$.
    }
    \label{fig:Wigner+marginals}
\end{figure*}

\subsection{Numerical optimization}
\label{sec:numoptimization}

The optimal pulses that implement target bosonic states are numerically optimized using Q-CTRL's graph-based optimizer (Boulder Opal \cite{boulder_opal1, boulder_opal2}). We select a model-based optimization procedure wherein the target Hamiltonian is encoded as a computational graph; each output pulse is represented by a complex-valued piecewise-constant function, where each segment is an optimizable variable. The Fock space of the bosonic modes is chosen such that truncation of the Hilbert space leads to infidelities below $8 \times 10^{-3}$. 

We select a gradient-based numerical optimizer that minimizes the cost function, $\mathcal{C}$, encoded as a graph node,
\begin{equation}
    \label{eq:cost_function}
    \mathcal{C} = (1-\mathcal{F}_\mathrm{th.}) + \epsilon \frac{T}{T_\mathrm{max}},
\end{equation}
where $\mathcal{F}_\mathrm{th.} = |\bra{\downarrow, \psi_{\mathrm{target}}} \hat{U} \ket{\downarrow, 0}|^2$ is the state fidelity. The ideal qubit-bosonic state is given by $\ket{\downarrow, \psi_{\mathrm{target}}}$ and the unitary evolution of the state preparation pulse is given by $\hat{U} = e^{-i \int^T_0 \hat{H}(t) dt}$ where $\hat{H}(t)$ is the control Hamiltonian of Eq.~1. $T_\mathrm{max} = 2.4 \textrm{ ms}$ is the maximum allowed duration; $\epsilon$, set to $0.05$, is chosen such that the optimizer minimises the duration while ensuring that $1-\mathcal{F}_\mathrm{th.} \lesssim \epsilon$. 

The pulses are constructed to be robust against motional dephasing by considering the Hamiltonian
\begin{equation}
    \hat{H}_k(t) = \hat{H}(t) + \delta_k \hat{a}^\dagger \hat{a},
\end{equation}
where $\delta_k$ is an error term that describes motional frequency offsets. The pulses are made robust by minimizing a total cost function that is averaged over several realizations of $\hat{H}_k(t)$, each with different error terms $\delta_k$. In practice, the frequency offsets are chosen from $\delta_k \in \{-k \sigma , \cdots ,k\sigma\}$, where $k \in \mathbb{Z}$ and $\sigma$ are akin to the variance of the noise. The total cost is computed as a weighted average of the infidelities obtained for each noise realization, where the weights correspond to a Gaussian envelope. With this, the cost function of Eq.~\ref{eq:cost_function} becomes
\begin{equation}
    C = \sum_{k=-n}^{n} e^{-k^2} \left(1- |\bra{\downarrow,\psi_\mathrm{target}} \hat{U}_k\ket{\downarrow,0}|^2 \right)+ \epsilon \frac{T}{T_\mathrm{max}},
\end{equation}
where $\hat{U}_k = e^{-i \int^T_0 \hat{H}_k(t) dt}$.

We impose various constraints on the pulses, which are captured as nodes in the optimization graph, in order to make them experimentally friendly to implement. First, only the phases $\phi_r$ and $\phi_b$ of the red- and blue-sideband fields can be varied. We set the amplitudes of the red- and blue-sideband fields to be constant throughout the pulse to avoid time-varying Stark shifts. We then limit the slew rates of the phases and add filtering in the form of a cutoff frequency to avoid signal distortions from the finite bandwidth of the AOM. The filtering process occurs mid-optimization; first, we generate a slew rate constrained pulse with $N_\mathrm{seg} = 90$ optimizable segments, we then apply a sinc filter and resample to retrieve a new pulse with $N_\mathrm{seg} = 240$ segments. We heuristically set the slew rate $\mathrm{SR} =\max \left|d \phi_{r,b}(t) / d t \right|$ to $\mathrm{SR} \times T = 2\pi \times 267$, where $T$ is the total pulse duration. The cutoff frequency $f_c \times T  = 2\pi \times 15$ is chosen to be much smaller than the AOM's bandwidth. 

Before implementing the pulse in the experiment, the discrete red- and blue-sideband phases are interpolated with cubic splines to remove discontinuities in the waveform. The waveform is implemented with an arbitrary waveform generator (Tabor Proteus P1284M) with a sampling clock of $\SI{1.25}{GSa/s}$. The number of segments in the pulse is sufficiently large such that the change in fidelity due to interpolation is negligible.

\subsection{Density matrix reconstruction}
\label{sec:dm_reconstruction}

The density matrix is determined from the experimentally reconstructed characteristic function using a convex optimizer~\cite{diamond2016cvxpy, Ahmed2021, AhmedGithub2021, Strandberg2022}. The optimizer finds a density matrix $\hat{\rho}$ (dimension $N\times N$) by minimizing the cost function
\begin{equation}
    \mathcal{C} = || \chi_\mathrm{opt}(\bm{\beta}) - \chi_\mathrm{exp}(\bm{\beta}) ||,
\end{equation}
where $\chi_\mathrm{opt}(\bm{\beta}) = \textrm{Tr}(\hat{\rho} \hat{\mathcal{D}}(\bm{\beta}))$, $\bm{\beta}$ is a vector of phase space points and $\chi_\mathrm{exp}(\beta)$ is the experimentally measured characteristic function. The convex optimizer is subject to the constraints $\textrm{Tr}(\hat{\rho}) = 1$ and $\hat{\rho} > 0$. To ensure the accuracy of this protocol, we increase the dimension of the Hilbert space used by the convex optimizer until the fidelity obtained from the reconstructed density matrix varies by less than $10^{-3}$. This corresponds to $N = 43$ for the GKP and binomial states, and $N = 60$ for the squeezed state.

We use bootstrapping to determine the uncertainties of fidelities and squeezing parameters computed from the reconstructed density matrices. For each reconstructed density matrix, we generate 100 synthetic characteristic function datasets. Each point of the characteristic function is calculated from repeated state-dependent fluorescence measurements, where $\mathrm{Re}[\chi(\beta)] = \langle \hat{\sigma}_z\rangle$. The probability $\langle \hat{\sigma}_z\rangle = M_B / M_T$ is calculated from the number of bright measurements, $M_B$, and the total number of measurements, $M_T$. The number of bright measurements is obtained by comparing the number of collected photons with a pre-determined threshold. Bootstrapping is done by randomly sampling 400 of the $M_T$ photon measurement outcomes for each point of the characteristic function, and recalculating $\langle \hat{\sigma}_z \rangle$. The convex optimizer is run on a High-Performance Computing (HPC) cluster that leverages parallelization.

\subsection{GKP state analysis}
\label{gkp_analysis}

\subsubsection{GKP codes}

We experimentally prepare two sub-classes of approximate GKP code words, the square (s) and hexagonal (h) states, which are defined as~\cite{Tzitrin2020,Grimsmo2021}
\begin{subequations}
\begin{alignat}{2}
 \ket{\mu}_{\mathrm{L}, \mathrm{s}.} &=\sum_{k=-\infty}^{\infty} & & e^{-\Delta^2|\alpha_{\mathrm{s}} (2k + \mu)|^2} \times \nonumber \\ 
 & & & \hat{\mathcal{D}}((2k + \mu) \alpha_{\mathrm{s}} ) \hat{S}(r) \ket{0}, \label{eq:gkp_definition} \\
\ket{\mu}_{\mathrm{L}, \mathrm{ h}.} & =\sum_{k, l=-\infty}^{\infty} & & e^{-\Delta^2|(2 k+\mu) \alpha_{\mathrm{h}}+l \beta_{\mathrm{h}}|^2} e^{-i \pi (k l + \frac{\mu l}{2})} \times \nonumber \\
& & & \hat{\mathcal{D}}((2 k + \mu) \alpha_{\mathrm{h}}+l \beta_{\mathrm{h}})|0\rangle,
\end{alignat}
\end{subequations}
where $\mu=0, 1$, $\alpha_\mathrm{s} = \sqrt{ \pi/2 }, \beta_\mathrm{s} =i\alpha_\mathrm{s}$ for the square state and $\alpha_\mathrm{h} = e^{-i\pi/3}\beta_\mathrm{h}$, $\beta_\textrm{h} = i\sqrt{\pi/\sqrt{3}}$ for the hexagonal state. The squeezing parameter $r = \ln{\Delta^{-1}}$ is related to the envelope parameter $\Delta \in [0,1]$.

\subsubsection{Wigner functions}

The Wigner functions of the GKP states are calculated from their reconstructed density matrices, using the relation \cite{Haroche2006}
\begin{equation}
    W(x, p) = \frac{1}{\pi} \int^{+\infty}_{-\infty} \bra{x-y} \hat{\rho} \ket{x+y} e^{2 i p  y} dy,
\end{equation}
where $x$, $p$ are the position and momentum variables associated with the operators $\hat{x}=(\hat{a}^{\dagger}+\hat{a}) / \sqrt{2}$ and $\hat{p}=i(\hat{a}^{\dagger}-\hat{a}) / \sqrt{2}$. The results are compared to theory in Fig.~\ref{fig:Wigner+marginals}, along with the corresponding marginal position and momentum probability distributions, $P(x) = \textrm{Tr}(\hat{\rho} \ket{x}\bra{x})$ and $P(p) = \textrm{Tr}(\hat{\rho} \ket{p}\bra{p})$, respectively.

\subsubsection{Squeezing analysis}

We analyze the envelope parameters of the GKP states, $\Delta$, using two methods. 

First, $\Delta$ can be calculated from the expectation of the stabilizer with the reconstructed density matrix \cite{Duivenvoorden2017, Flhmann2019},
\begin{equation}
\begin{aligned}
\Delta_X & =\frac{1}{2|\alpha|} \sqrt{-\log \left(\left|\operatorname{Tr}\left[\hat{S}_X \hat{\rho}\right]\right|^2\right)}, \\
\Delta_Z & =\frac{1}{2|\beta|} \sqrt{-\log \left(\left|\operatorname{Tr}\left[\hat{S}_Z \hat{\rho}\right]\right|^2\right)},
\end{aligned}
\end{equation}
where $\alpha,\beta$ are defined in the main text for the square or hexagonal code. The GKP squeezing parameter is given in decibels using $-10 \log_{10}(\Delta^2)$. The results from this method, reported in Table I, are \{5.5(2), 7.5(2), 6.5(3)\} for $\hat{S}_X$ and \{6.3(2), 7.5(2), 6.3(4)\} for $\hat{S}_Z$, corresponding to the states \{GKP$^{s}_{\Delta=0.247}$, GKP$^{s}_{\Delta=0.301}$, GKP$^{h}_{\Delta=0.301}$\}. The expectation values of the stabilizers can also be obtained from the characteristic function measurements, where $\langle \hat{S}_Z \rangle = \chi \left(\sqrt{2\pi} \right)$, $\langle \hat{S}_X \rangle = \chi \left(i \sqrt{2\pi} \right)$ for the square GKP state and $\langle \hat{S}_Z \rangle = \chi \Big(2 \sqrt{\pi/\sqrt{3}} + i \sqrt{\sqrt{3}\pi}/2 \Big)$, $\langle \hat{S}_X \rangle = \chi \Big(i \sqrt{\sqrt{3}\pi} \Big)$ for the hexagonal GKP state. From the data shown in Fig.~2, the squeezings of the states are found to be $\{5.04, 6.78, 7.06\}$~dB for $\hat{S}_X$ and $\{6.54, 8.45, 5.97\}$~dB for $\hat{S}_Z$.

A second method is reported in Ref. \cite{Eickbusch2022}, which we briefly summarize here. The effective squeezing of an experimentally prepared state $\hat{\rho}_\mathrm{exp}$ is determined by the value of $\Delta$ that maximizes $\mathrm{Tr}(\hat{P}_\Delta \hat{\rho}_\mathrm{exp})$, where $\hat{P}_\Delta = \ket{+Z_\Delta}\bra{+Z_\Delta} + \ket{-Z_\Delta}\bra{-Z_\Delta}$ is a projector onto the finite-energy GKP codespace. The $\ket{+Z_\Delta}$ eigenstate is determined by finding the ground state of the fictitious Hamiltonian $H = -\hat{S}_{X,\Delta} - \hat{S}_{Z, \Delta} - \hat{Z}_{\Delta}$, where $\hat{S}_{X,\Delta} = \hat{E}_\Delta \hat{\mathcal{D}}(2\alpha) \hat{E}^{-1}_\Delta$ and $\hat{S}_{Z,\Delta} = \hat{E}_\Delta \hat{\mathcal{D}}(2\beta) \hat{E}^{-1}_\Delta$ are the finite-energy stabilizers. $\hat{Z}_{\Delta} = \hat{E}_\Delta \hat{\mathcal{D}}(\beta) \hat{E}^{-1}_\Delta$ is the finite-energy Pauli-Z operator, with the envelope operator $\hat{E}_\Delta = e^{- \Delta^2 \hat{a}^\dagger \hat{a}}$. The $\ket{-Z_\Delta}$ eigenstate is found by applying the finite-energy Pauli-X operator $\hat{X}_\Delta = \hat{E}_\Delta \hat{\mathcal{D}}(\alpha) \hat{E}_\Delta^{-1}$. The GKP squeezing parameters obtained with this method are \{GKP$^{s}_{\Delta=0.247}$, GKP$^{s}_{\Delta=0.301}$, GKP$^{h}_{\Delta=0.301}$\} = \{10.4(1), 9.47(8), 9.1(1)\}, which are reported in the main text.

\subsubsection{Logical fidelity}

While the squeezing quantifies how close the GKP state is to the ideal GKP codespace, the logical fidelity quantifies the amount of logical information contained in the state~\cite{Kolesnikow2024}, which we compute as $\bar{\mathcal{F}} = \bra{Z_+} D(\hat{\rho}_\mathrm{exp})\ket{Z_+}$, where $\ket{Z_+}$ is the $+1$ eigenstate of the $2\times 2$ Pauli-$Z$ operator. Here, $D(\cdot)$ is a decoder that transforms the bosonic state into a logical $2\times 2$ density matrix that encodes the logical information. 
	
To obtain the decoded density matrix, we utilize the Stabiliser Subsystem Decomposition introduced in Ref.~\cite{Shaw2024}, together with an approach recently developed in Ref.~\cite{Shaw2024a} and used in Ref.~\cite{Kolesnikow2024}. Here, the decoded density matrix is expressed in the Pauli basis as
\begin{equation}
    D(\hat{\rho}_\text{exp}) = \frac{1}{2} \left( \hat{I} + \langle \hat{X}_\text{m} \rangle \hat{X} + \langle \hat{Y}_\text{m} \rangle \hat{Y} + \langle \hat{Z}_\text{m} \rangle \hat{Z} \right), \label{eq:decoded_dm}
\end{equation}
where $\{ \hat{I}, \hat{X}, \hat{Y}, \hat{Z} \}$ are the usual $2 \times 2$ Pauli operators, and $\{ \hat{X}_\text{m}, \hat{Y}_\text{m}, \hat{Z}_\text{m} \}$ are the Pauli \textit{measurement} operators, which are Hermitian operators that act on the full Hilbert space for the bosonic mode. The Pauli measurement operators are defined in such a way that the decoded density matrix corresponds to the state obtained by averaging over many rounds of ideal GKP error correction~\cite{Shaw2024}. Expressions for these operators in the displacement basis for the square and hexagonal GKP code are calculated in \cite{Shaw2024a}, which facilitate a way to experimentally obtain the expectation values in Eq.~\ref{eq:decoded_dm} from a subset of the characteristic function via
\begin{subequations}
\begin{align}
    \langle \hat{X}_\text{m} \rangle &= \frac{1}{\pi} \sum_{n = -\infty}^\infty \frac{(-1)^n}{n+1/2} \; \chi\left(\sqrt{2\pi}(n+1/2)\right), \\
    \langle \hat{Y}_\text{m} \rangle &= \frac{1}{\pi^2} \sum_{m,n = -\infty}^\infty \frac{\chi\left(\sqrt{2\pi}(m+1/2) + i\sqrt{2\pi}(n+1/2)\right)}{(m+1/2)(n+1/2)}, \\
    \langle \hat{Z}_\text{m} \rangle &= \frac{1}{\pi} \sum_{n = -\infty}^\infty \frac{(-1)^n}{n+1/2} \; \chi\left(i\sqrt{2\pi}(n+1/2)\right),
\end{align}
\end{subequations}
for the square GKP code and
\begin{widetext}
\begin{subequations}
    \begin{align}
        \langle \hat{X}_\text{m} \rangle &= \frac{3}{\pi^2} \sum_{m,n = -\infty}^\infty f(m,n) \; \chi\left((m+1/2)\sqrt{3}|\alpha_\text{h}| + i(m + 2n + 1/2)|\alpha_\text{h}|\right), \\
        \langle \hat{Y}_\text{m} \rangle &= \frac{3}{\pi^2} \sum_{m,n = -\infty}^\infty f(n,n-m) \; \chi\left((m+1/2)\sqrt{3}|\alpha_\text{h}| + i(m + 2n + 3/2)|\alpha_\text{h}|\right),  \\
        \langle \hat{Z}_\text{m} \rangle &= \frac{3}{\pi^2} \sum_{m,n = -\infty}^\infty f(n,m) \; \chi\left(m\sqrt{3}|\alpha_\text{h}| + i(m + 2n + 1)|\alpha_\text{h}|\right),
    \end{align}
\end{subequations}
\end{widetext}
for the hexagonal GKP code, where $f(m,n) = \frac{(-1)^m \cos\left( \frac{\pi}{3}(m + n - 1) \right)}{(m+n+\frac{1}{2})(m - 2n + \frac{1}{2})}$. For approximate GKP states, the infinite sums in these expressions may be truncated such that the characteristic function has negligible support at points in phase space beyond this truncation. For the states considered in this Letter, we use a truncation of $\pm 3$ (including the left end-point but not the right), corresponding to the largest points in phase space used to construct the full characteristic functions for the states shown in Fig. 2 of the main text. This requires a total of 48 and 108 (15 and 27 after accounting for symmetry) measurements per shot for the square and hexagonal GKP states, respectively, substantially less than the 2401 (625 after accounting for symmetry) measurements per shot used to construct the full characteristic function. Together with the squeezing parameter, which requires only two measurements, this provides a streamlined approach to characterize a GKP state using far fewer measurements than are required for full state tomography.

\subsubsection{Comparison with gate-based approaches}

We compare our pulse-level optimization protocol with gate-based approaches for the deterministic creation of GKP states. We restrict ourselves to trapped-ion implementations. 

The proposal outlined in Ref.~\cite{Hastrup2021} for the deterministic preparation of GKP states is briefly summarized in the following. 
The protocol sequentially applies state-dependent forces after initializing the bosonic mode to a squeezed state $\ket{r} = \hat{S}(r) \ket{0}$. The total pulse sequence is $\prod_k^N e^{i u_k \hat{\sigma}_x\hat{x}}e^{i w_k \hat{\sigma}_y \hat{x}}e^{i v_k \hat{\sigma}_x \hat{p}}$, where the parameters $u_k$, $w_k$ and $v_k$ are determined from analytical formulae. The quality of the GKP state improves with increasing $N$. The total duration is determined from the duration of the squeezing operation, $T_s$, and the duration of the state-dependent forces, $T_\mathrm{SDF}$. We assume that squeezing is enacted by second-order sideband interactions, where the duration is $T_s = r/(\eta^2\Omega)$ \cite{Kienzler2015}. The duration of the SDF pulses is determined by the sum of their displacement magnitudes, $T_\textrm{SDF} = \sum_k^N 2(|u_k| + |w_k| + |v_k|)/(\eta\Omega)$.  

\begin{figure}
    \centering
    \includegraphics[width=8cm]{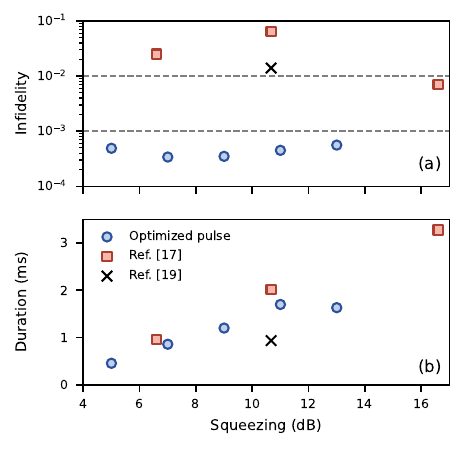}
    \caption{\textbf{Comparison of GKP state preparation with optimized pulses (this work) against gate-based approaches.} 
    \textbf{(a)} Numerically calculated infidelities and \textbf{(b)} total pulse/sequence durations are plotted with respect to target GKP squeezing parameters. Dashed lines indicate infidelities of $10^{-3}$ and $10^{-2}$ to guide the eye. The method of Ref.~\cite{Hastrup2021} (red squares) consists of applying $N$ rounds of sequential state-dependent forces, where more rounds increase the quality of the states. Here, we plot the performance up to $N=3$ rounds. Similarly, Ref.~\cite{Neeve2022} (black crosses) also considers sequential application of state-dependent forces. }
    \label{fig:gkp_scaling}
\end{figure}

A similar method is demonstrated in Ref.~\cite{Neeve2022}, where the following sequence is applied to create a GKP $\ket{1}_\mathrm{L}$ logical state,

\begin{equation}
    \ket{\downarrow}\otimes\ket{1}_\mathrm{L}  = e^{-i \alpha_4 \hat{\sigma}_y \hat{x} } e^{i \alpha_3 \hat{\sigma}_x \hat{p} } e^{i \alpha_2 \hat{\sigma}_y \hat{x}}   e^{-i \alpha_1 \hat{\sigma}_x \hat{p} } \ket{\downarrow}\otimes \ket{r} .
\end{equation}

The displacement magnitudes are $\alpha_1 = \alpha_\mathrm{grid}$, $\alpha_2 = 0.031 \alpha_\mathrm{grid}$, $\alpha_3 = 0.5 \alpha_\mathrm{grid}$, $\alpha_4 = 0.125 \alpha_\mathrm{grid}$ and $\alpha_\mathrm{grid} = 2 \sqrt{\pi}$. Note that this preparation method closely resembles the scheme proposed in Ref.~\cite{Hastrup2021} for $N=2$. The squeezing operation was implemented with reservoir engineering, which has a negligible duration compared to the SDF pulse durations, $T_\textrm{SDF} = 2 \sum_k |\alpha_k| / (\eta\Omega)$. 

We investigate the trade-off between the infidelity and total duration with a given set of target GKP squeezing parameters (see Fig.~\ref{fig:gkp_scaling}). We target a GKP $\ket{1}_\mathrm{L}$ state by setting $\mu=1$ in Eq.~\ref{eq:gkp_definition}. Infidelities are computed by numerically simulating the SDF pulses and using the fidelity equation defined in Table I. Our results were obtained by setting $\epsilon = 10^{-2}$ in the cost function of Eq.~\ref{eq:cost_function} to minimize the duration of the pulses while retaining high fidelities. The amplitudes of the red- and blue-sideband fields are kept constant. However, we remove the slew rate and frequency filtering constraints on the phases to increase the flexibility of the optimizer. We also reduced the number of segments in the range $N_\mathrm{seg} \in [70, 90]$. The Rabi frequency of the first order sideband is set to $\eta \Omega = 2\pi \times \SI{2}{kHz}$ and we set the Lamb-Dicke parameter $\eta=0.1$.

Figure~\ref{fig:gkp_scaling} shows that optimized pulses achieve better fidelities for a wide range of target squeezings while maintaining durations comparable to the alternative methods. Notably, our method does not require initializing the bosonic state to a squeezed state. Moreover, our numerical optimizer is more flexible than the gate-based approaches in that the target squeezings are not constrained to discrete values. Figure~\ref{fig:gkp_scaling} reveals a large gap between squeezing parameters with increasing N. One can also leverage this flexibility to explore trade-offs between the duration and the fidelity, i.e. pulses can be made faster at the expense of larger errors.

\subsection{Binomial states}
\label{binomial_states}

Binomial codes protect against errors that include up to $L$ boson losses, $\hat{a}^L$, $G$ boson gains, $(\hat{a}^\dagger)^G$, and $D$ dephasing events, $(\hat{a}^\dagger\hat{a})^D$~\cite{Marios2016}. The corresponding codewords are conveniently expressed using binomial coefficients, 

\begin{equation}
\left|{\pm Z}\right\rangle_\mathrm{bin}=\frac{1}{\sqrt{2^N}} \sum_{\text {p even/odd }}^{[0, N+1]} \sqrt{\begin{pmatrix} N+1 \\ p  \end{pmatrix}}|p(S+1)\rangle,
\end{equation}
where the spacing is $S=L+G$, the maximum order is $N=\max \{L, G, 2 D\}$, and the index $p$ ranges from 0 to $N+1$. If one only considers loss (gain) errors, the distance of the code is $L+1$ ($G+1$) and represents the minimum number of loss (gain) events needed to mix two code words. 

In the main text, we prepare the $+Z$ eigenstate of a codeword with spacing $S=1$ and order $N=1$, resulting in $\ket{+Z_{S=1}}_\mathrm{bin} = \frac{1}{\sqrt{2}}(\ket{0} + \ket{4})$. We also prepare the $\pm Z$ eigenstates of a codeword with spacing $S=2$ and order $N=2$, resulting in $\ket{+Z_{S=2}}_\mathrm{bin} = \frac{1}{2}(\ket{0} + \sqrt{3}\ket{6})$ and $\ket{-Z_{S=2}}_\mathrm{bin} = \frac{1}{2}(\sqrt{3}\ket{3} + \ket{9})$.

\subsection{Analysis of squeezed vacuum state}
\label{sec:squeezed_analysis}

\begin{figure}
    \centering
    \includegraphics[width=8cm]{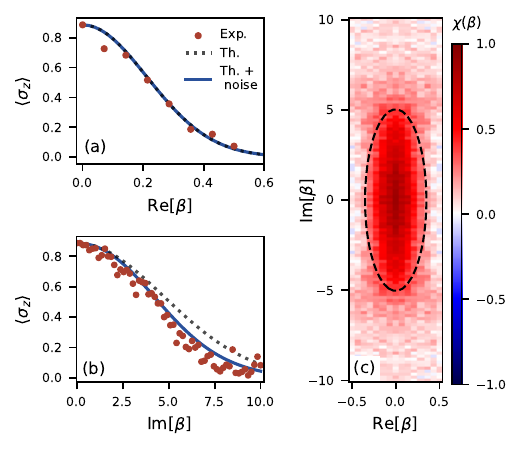}
    \caption{\textbf{Squeezed state analysis.} 
    \textbf{(a, b)} One-dimensional slices of the characteristic function along the squeezed ($\mathrm{Re}[\beta]$) and anti-squeezed ($\mathrm{Im}[\beta]$) axes. 
    Theory data (dotted line) corresponds to an ideal squeezed state with $r=1.55$. We also plot numerically simulated one-dimensional slices that include motional dephasing noise during the reconstruction protocol (solid blue line). Theory lines are scaled by the experimentally measured value at $\chi(\beta=0)$.
    \textbf{(c)} Two-dimensional characteristic function, reproduced from Fig.~2. A fit to Eq.~\ref{Eq.squeezeanalytic} (dashed line) yields the squeezing parameter.
    }
    \label{fig:squeezed_state_analysis}
\end{figure}

We analyze the experimentally created squeezed state (Fig.~2(a)) with two methods previously used in the literature. First, the squeezing parameter is computed from the reconstructed density matrix by fitting the probability distribution $P(x) = \textrm{Tr}(\hat{\rho} \ket{x}\bra{x})$ to a Gaussian function and extracting the variance~\cite{Eickbusch2022}. We find that estimating a squeezing parameter with this method can be heavily skewed due to decoherence during state reconstruction. Due to the large extent of the characteristic function in phase space, the anti-squeezed quadrature suffers from decoherence, which corrupts the density matrix reconstruction. We verify this in Fig.~\ref{fig:squeezed_state_analysis} by numerically simulating the noisy reconstruction of one-dimensional slices of the characteristic function. We perform a master equation simulation on an ideal squeezed state with $r=1.55$. We include a Linbladian dephasing term $\sqrt{\Gamma} \hat{a}^\dagger \hat{a}$ with $\Gamma = \SI{18}{Hz}$ during the displacement pulses, and a constant offset $\delta\hat{a}^\dagger \hat{a}$ with $\delta/2\pi = \SI{18}{Hz}$ to model the effects of drift. We find that, for large values of $|\beta|$, the effects of dephasing become prominent, and the noise model is in good agreement with the experiment.

To investigate the effect of dephasing during reconstruction on the squeezing estimation, we extend the above numerical simulations to generate a noisy reconstructed two-dimensional characteristic function. The density matrix is determined with the convex optimizer, from which we find a fidelity of 0.906 and a squeezing of 11.17 dB instead of 13.46 dB. This indicates that the density matrix reconstruction method is sensitive to dephasing when applying the $\hat{\mathcal{D}}(\pm\beta/2)$ pulse during state reconstruction. The value of $\Gamma$ is consistent with the motional coherence time measured in a separate experiment. Applying this method to the density matrix obtained from the experimental results of Fig.~\ref{fig:gkp_scaling}, we obtain a squeezing of $\SI{11.52}{dB}$.

Alternatively, the squeezing parameter $r$ can be extracted from the experimental characteristic function measurements \cite{Flhmann2019}. As pointed out in Ref.~\cite{Hastrup2021sq}, this method reduces errors from overestimating the squeezing from indirect measurements, which are sensitive to small fluctuations and coherence in the state. The squeezing is extracted from $\chi(\beta)$ by fitting the data to
\begin{equation}
    \label{eq:chi_func_squeezed}
    P(\langle \hat{\sigma}_z \rangle) = A \chi_S(\beta),
\end{equation}
where $A$ is a scalar to take into account the state preparation and measurement (SPAM) error due to residual spin-motion entanglement, and the analytical characteristic function of a squeezed state is
\begin{equation}
    \chi_S(\beta) = e^{-|\beta \cosh(r) + \beta^* \sinh(r)|^2/2}.
    \label{Eq.squeezeanalytic}
\end{equation}

\begin{figure*}
    \centering
    \includegraphics[]{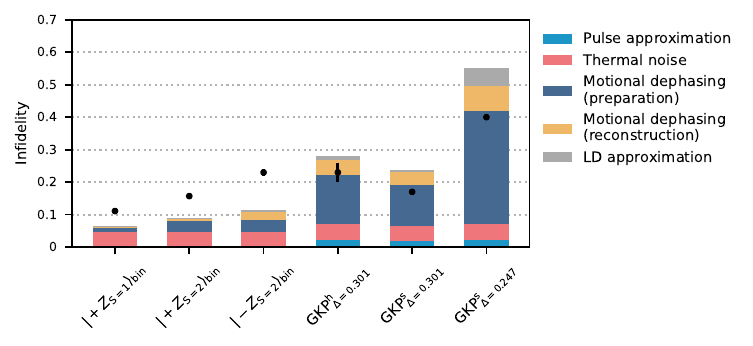}
    \caption{\textbf{Error budget for the bosonic states created in this work.} 
    Predicted infidelities (bars) are plotted alongside the experimentally measured infidelities (black circles) reported in Table I. The binomial state labeled $\ket{+Z_\mathrm{S=1}}_\mathrm{bin}$ corresponds to $(\ket{0} + \ket{4})/\sqrt{2}$, while the labels $\ket{+Z_\mathrm{S=2}}_\mathrm{bin}$ and $\ket{-Z_\mathrm{S=2}}_\mathrm{bin}$ correspond to $(\ket{0} + \sqrt{3}\ket{6})/2$ and $(\sqrt{3}\ket{3}+\ket{9})/2$ respectively. We find that thermal noise is an important error source for all states. We also find that motional dephasing significantly impacts the GKP states more than the binomial states. We observe that dephasing during the reconstruction is a non-negligible error source. Infidelities associated with the Lamb-Dicke approximation become important for GKP states with larger squeezings.}
    \label{fig:error_budget}
\end{figure*}

We again investigate the effects of dephasing during reconstruction and apply the same master equation simulations described above. Fitting Eq.~\ref{eq:chi_func_squeezed} to the simulated characteristic function, we find a squeezing of 12.14 dB instead of 13.46 dB. While this method also suffers from decoherence during the reconstruction, it gives a closer estimate than the density matrix method.

We fit the experimental data (Fig.~2(a)) to Eq.~\ref{eq:chi_func_squeezed} and fix $A = 0.887$, which is the experimentally measured value at $\chi_S(\beta=0)$. We extract $r = 1.487(5)$, which corresponds to a squeezing of 12.91(5) dB (see Fig.~\ref{fig:squeezed_state_analysis}(c)). We report this squeezing parameter as the representative value of the squeezed vacuum state, as this method avoids some of the inaccuracies of the density matrix estimation caused by decoherence during the reconstruction protocol. 

We also investigate the difference in squeezings along the squeezed and anti-squeezed axes by fitting 1-dimensional slices of data along $\textrm{Re}[\beta]$ and $\textrm{Im}[\beta]$ to Eq.~\ref{eq:chi_func_squeezed}. We extract values of $\SI{13.6(2)}{dB}$ and $\SI{-11.55(8)}{dB}$ along the squeezed and anti-squeezed axes, respectively.

\subsection{Error budget}
\label{sec:error_budget}

We build an error budget to identify the dominant noise mechanisms that limit the state fidelities reported in Table I. We consider five error sources: (i) theoretical imperfection of the control pulse to create target states, (ii) finite motional temperature, (iii) motional dephasing during preparation, (iv) motional dephasing during reconstruction, and (v) Lamb-Dicke approximations. The pulse infidelity is determined by how well the state created by the pulse approximates the target state. Finite motional temperature arises when the motional mode is not perfectly cooled to the ground state. This is predominantly due to imperfect sideband cooling and photon recoils. Motional dephasing arises from fluctuations of the motional mode frequency due to noise in the electronics and the resonator for the trapping radio frequency signal. We separately consider dephasing during preparation and reconstruction. Errors associated with the Lamb-Dicke approximation arise from higher-order terms in the interaction Hamiltonian that are neglected during pulse optimization and numerical simulations. We do not consider errors from motional heating, as the experimentally measured heating rate is negligible ($\dot{\bar{n}} = \SI{0.2}{s^{-1}}$).

The errors are computed by numerically simulating the pulse with the above imperfections. Infidelities associated with the control pulse are computed in the absence of decoherence. Finite temperature is simulated by considering a thermal state with $\bar{n} = 0.05$, which is consistent with measurements in separate experiments. We simulate motional dephasing with a Lindblad master equation, using the Linbladian dephasing term $\sqrt{\Gamma}\hat{a}^\dagger \hat{a}$ with $\Gamma = \SI{18}{Hz}$. We also include constant frequency offsets $\delta \hat{a}^\dagger \hat{a}$, with $\delta = 2\pi \times \SI{18}{Hz}$, to simulate the effects of slow drift in our system. Errors from pulse approximation, thermal noise, and motional dephasing during preparation are found by numerically simulating the pulse and calculating the fidelity as defined in Table I with respect to the ideal qubit-bosonic state. Errors during the reconstruction are estimated by retrieving the density matrix from the numerically simulated characteristic function and calculating the fidelity with respect to the ideal bosonic state. The resulting error budget is reported in Fig.~\ref{fig:error_budget}. 

Errors associated with the Lamb-Dicke approximation are estimated by considering the Hamiltonian,
\begin{equation} \label{eq:exact_sdf_hamiltonian}
    \hat{H}_\mathrm{SDF} = \hat{H}_\mathrm{int}(t, - \omega_r) + \hat{H}_\mathrm{int}(t, \omega_r),
\end{equation}
with
\begin{align} \label{eq:int_hamiltonian_exact}
    \hat{H}_\mathrm{int}(t, \Delta\omega_L) = \frac{\Omega}{2}\hat{\sigma}^+ & \mathrm{exp}\left(i \eta (\hat{a} e^{- i \omega_r t} + \hat{a}^\dagger e^{i \omega_r t})\right) \nonumber \\
    & \times  e^{- i \Delta \omega_L t} e^{i \phi(t)} + \mathrm{h.c.},
\end{align}
where $\eta$ is the Lamb-Dicke parameter, $\omega_r$ is the radial mode frequency, and $\Delta \omega_L$ is the laser's frequency detuning from the carrier transition. The approximate Hamiltonian of Eq.~1 is obtained by expanding the exponential of Eq.~\ref{eq:int_hamiltonian_exact}, $\mathrm{exp}\left(i \eta (\hat{a} e^{- i \omega_r t} + \hat{a}^\dagger e^{i \omega_r t})\right)$, with a Taylor series, then setting $\Omega_r = \Omega_b = \eta \Omega$, and finally dropping rapidly oscillating terms under a rotating wave approximation in the limit $\eta \Omega \ll \omega_r$. Errors may, therefore, appear when higher-order terms are not negligible.

Infidelities from the Lamb-Dicke approximation are calculated by numerically simulating the Hamiltonian of Eq.~\ref{eq:exact_sdf_hamiltonian} with up to $4^\mathrm{th}$ order terms in the Taylor expansion. The Lamb-Dicke parameter is $\eta = z_0 k = 0.083$, where $z_0 = \sqrt{\hbar / 2 m \omega_r}$ is the spatial extent of the motional waveform and $k = 2\pi/\lambda$ is the wavevector, with $\lambda = \SI{355}{nm}$.


\end{document}